\documentclass[aps,pra,twocolumn,longbibliography] {revtex4-1}
\usepackage{amsmath,graphicx,amsfonts}

\usepackage{times}
\usepackage[colorlinks=true,linkcolor=blue,urlcolor=blue,citecolor=blue, breaklinks=true,hypertexnames=false]{hyperref}

\usepackage{amssymb}
\usepackage{bbm}

\usepackage[utf8]{inputenc}
\DeclareUnicodeCharacter{03BC}{\mbox{$\mu$}}
\DeclareUnicodeCharacter{03B4}{$\delta$}
\DeclareUnicodeCharacter{2009}{-}

\usepackage[english]{babel}
\addto\captionsenglish{}

\makeatletter
\def\bbl@set@language#1{%
	\edef\languagename{%
		\ifnum\escapechar=\expandafter`\string#1\@empty
		\else\string#1\@empty\fi}%
	\@ifundefined{babel@language@alias@\languagename}{}{%
		\edef\languagename{\@nameuse{babel@language@alias@\languagename}}%
	}%
	\select@language{\languagename}%
	\expandafter\ifx\csname date\languagename\endcsname\relax\else
	\if@filesw
	\protected@write\@auxout{}{\string\select@language{\languagename}}%
	\bbl@for\bbl@tempa\BabelContentsFiles{%
		\addtocontents{\bbl@tempa}{\xstring\select@language{\languagename}}}%
	\bbl@usehooks{write}{}%
	\fi
	\fi}
\newcommand{\DeclareLanguageAlias}[2]{%
	\global\@namedef{babel@language@alias@#1}{#2}%
}
\makeatother

\DeclareLanguageAlias{en}{english}

\makeatletter
\def\@bibdataout@aps{%
	\immediate\write\@bibdataout{%
		@CONTROL{%
			apsrev41Control%
			\longbibliography@sw{%
				,author="08",editor="1",pages="1",title="0",year="1"%
			}{%
				,author="08",editor="1",pages="1",title="",year="1"%
			}%
		}%
	}%
	\if@filesw \immediate \write \@auxout {\string \citation {apsrev41Control}}\fi 
}
\makeatother

\date\today
\begin{document}
\title{Exact result for the polaron mass in a one-dimensional Bose gas}
\author{Zoran Ristivojevic}
\affiliation{Laboratoire de Physique Th\'{e}orique, Universit\'{e} de Toulouse, CNRS, UPS, 31062 Toulouse, France}
	
\begin{abstract}
We study the polaron quasiparticle in a one-dimensional Bose gas. In the integrable case described by the Yang-Gaudin model, we derive an exact result for the polaron mass in the thermodynamic limit. It is expressed in terms of the derivative with respect to the density of the ground-state energy per particle of the Bose gas without the polaron. This enables us to find high-order power series for the polaron mass in the regimes of weak and strong interaction. 
\end{abstract}
\maketitle

Polarons were originally introduced to represent electrons that had changed their properties due to the influence of lattice distortions in ionic crystals \cite{pekar_1963}. The motion of electrons in such environments can be strongly modified as they move together with a surrounding cloud of phonons. This gives rise to a polaron mass that is larger than the electron one \cite{landau_effective_1948}. The concept of polarons is nowadays wider and applies well beyond solids, with particular importance for ultracold gases. Such environments offer numerous possibilities where polarons can be realized. Experiments provided both higher-dimensional  \cite{schirotzek_observation_2009,koschorreck_attractive_2012,kohstall_metastability_2012,jorgensen_observation_2016,hu_bose_2016,yan_bose_2020,skou_non-equilibrium_2021} and one-dimensional \cite{catani_quantum_2012,meinert_bloch_2017} systems which host polarons.

Many theoretical papers addressed polaron properties in one-dimensional  bosonic systems \cite{li_exact_2003,astrakharchik_motion_2004,fuchs_spin_2005,batchelor_collective_2006,sacha_self-localized_2006,kamenev_dynamics_2009,zvonarev_edge_2009,schecter_dynamics_2012,mathy_quantum_2012,burovski_momentum_2014,petkovic_dynamics_2016,robinson_motion_2016,grusdt_bose_2017,parisi_quantum_2017,volosniev_analytical_2017,robinson_excitations_2017,kain_analytical_2018,panochko_mean-field_2019,petkovic_microscopic_2020,jager_strong-coupling_2020}. Several of them \cite{li_exact_2003,fuchs_spin_2005,batchelor_collective_2006,zvonarev_edge_2009,robinson_motion_2016,robinson_excitations_2017} rely on models solvable by the Bethe ansatz \cite{gaudin_systeme_1967,yang_exact_1967,sutherland_further_1968}.  Theoretical studies of such models are important for experiments, which are amenable to simulate them. Moreover, studies of integrable models can give rise to exact analytical results for physically relevant quantities, which are \textit{per se}  important.

A one-dimensional system of bosons with contact interaction is described by the Hamiltonian
\begin{align}\label{eq:H}
H=\frac{\hbar^2}{2m}\left[-\sum_{j=1}^{N} \frac{\partial^2}{\partial x_j^2}+c\sum_{j\neq l} \delta(x_j-x_l)\right]. 
\end{align}
Here $m$ is the particle mass, $c>0$ controls the interaction strength, and $N$ is the total number of particles in the system. We consider the case with periodic boundary conditions. For a single-component Bose gas, the Hamiltonian (\ref{eq:H}) is exactly solved by the Bethe ansatz technique and is known as the Lieb-Liniger model  \cite{lieb_exact_1963,lieb_exact2_1963}.

The Hamiltonian (\ref{eq:H}) is solvable by Bethe ansatz for more general  symmetries of the wave function \cite{yang_exact_1967,sutherland_further_1968,gaudin_systeme_1967}. Let us consider a two-component (isospin-$\frac{1}{2}$) Bose gas \cite{gaudin_systeme_1967,yang_exact_1967,li_exact_2003}. In this case the Hamiltonian (\ref{eq:H}) is known as the Yang-Gaudin model for a Bose gas. Its eigenstates can be classified with respect to the value of the total isospin. In the sector where it has the maximal value $N/2$, one studies a single-component system described by the Lieb-Liniger model. It is characterized by $N$ density quantum numbers $I_1,I_2,\ldots, I_N$ that define $N$ quasimomenta, $k_1,k_2,\ldots, k_N$. In the case of the total isospin $N/2-1$, which will be the focus of this paper, the system acquires an additional spin quantum number that defines spin rapidity $\eta$. The corresponding Bethe ansatz equations can be expressed as \cite{zvonarev_edge_2009,Note1}\footnotetext{Note that Eq.~(\ref{eq:BE}) conveniently contains an extra additive term $\pi$ and thus it has different quantum numbers $I_j$ from those in Refs.~\cite{li_exact_2003,fuchs_spin_2005,robinson_excitations_2017}.}
\begin{align}\label{eq:BE}
k_j L+\sum_{l=1}^N \theta(k_j-k_l)={2\pi}I_j+\pi+\theta(2k_j-2\eta),
\end{align}
where $L$ is the system size, $I_j$ are integers for odd $N$ and odd half-integers for even $N$, and $\theta(k)=2\arctan(k/c)$ is, up to the sign, the scattering phase shift. The momentum $p$ and the energy $E$ of the system are
\begin{align}\label{eq:EP}
 p=\hbar \sum_{j=1}^N k_j,\quad E=\frac{\hbar^2}{2m}\sum_{j=1}^N k_j^2.
\end{align}
Note that the spin rapidity $\eta$ indirectly enters  Eq.~(\ref{eq:EP}) through the quasimomenta $k_j$ that depend on $\eta$ via Eq.~(\ref{eq:BE}).

In the special case $\eta\to+\infty$, the Bethe ansatz equations (\ref{eq:BE}) describe the quasimomenta of the single-component (i.e., isospin-polarized) system described by the Lieb-Liniger model \cite{lieb_exact_1963}. Its ground state is realized for a symmetric configuration of quasimomenta around zero, which occurs for
\begin{align}\label{eq:Ij}
I_j=j-\frac{N+1}{2},\quad j=1,2,\ldots,N.
\end{align}
The density of quasimomenta $\rho(k_j)=[L(k_{j+1}-k_j)]^{-1}$ in the thermodynamic limit ($N\to\infty$ and $L\to\infty$ such that the density $n=N/L$ is fixed) satisfies the integral equation \cite{lieb_exact_1963}
\begin{align}\label{eq:rho}
\rho(k,Q)-\frac{1}{2\pi}\int_{-Q}^{Q} dq\:\!\theta'(k-q)\rho(q,Q)=\frac{1}{2\pi}.
\end{align}
Here $\theta'(k)=d\theta(k)/dk$ and $Q$ is the Fermi rapidity, which denotes the largest occupied quasimomentum in the ground state. The density of quasimomenta enables us to express the particle density as
\begin{align}\label{eq:n}
n(Q)=\int_{-Q}^{Q} dk\:\!\rho(k,Q),
\end{align}
where we emphasized that $n$ is a function of $Q$. The ground-state momentum is zero and the ground-state energy per particle $\epsilon_0$ is given by \cite{lieb_exact_1963}
\begin{align}\label{eq:E0}
\epsilon_0=\frac{\hbar^2}{2m\:\!n}\int_{-Q}^{Q} dk\:\!k^2\rho(k,Q).
\end{align}
We note that the Fermi rapidity $Q$ naturally enters $\epsilon_0$ in Eq.~(\ref{eq:E0}). However we can eventually express $Q$ in terms of $n$ by inverting their connection (\ref{eq:n}).

At finite $\eta$, Eq.~(\ref{eq:BE}) describes the Bose gas with one isospin reversed. In the case of an unchanged set of quantum numbers (\ref{eq:Ij}), the system is in an excited state that is called spin-wave excitation \cite{fuchs_spin_2005} or a magnon \cite{zvonarev_edge_2009}. In the context of this paper, such collective excitation of the system describes a polaron quasiparticle excitation. 

The momentum of the system in the excited state coincides with the momentum of the polaron excitation. Substituting $k_j$ expressed from Eq.~(\ref{eq:BE}) into  Eq.~(\ref{eq:EP}), in the thermodynamic limit we obtain \cite{zvonarev_edge_2009}
\begin{align}\label{eq:p}
p(Q,\eta)=\hbar \int_{-Q}^{Q} dk\:\! \rho(k,Q)\left[\pi+\theta(2k-2\eta)\right].
\end{align}
We note that the momentum explicitly depends on the Fermi rapidity $Q$ and the spin rapidity $\eta$. As expected, at $\eta\to+\infty$ we obtain $p=0$, i.e., the polaron is not excited. 

The evaluation of the energy of the system in the excited state is more involved. At finite $\eta$, the quasimomenta become shifted by $\Delta k_j=k_j(\eta)-k_j(\eta\to+\infty)=O(1/L)$. From Eq.~(\ref{eq:BE}) we then find
\begin{align}\label{eq:Deltak}
	\Delta k_j L={}&-\sum_{l=1}^{N}\theta'(k_j-k_l)(\Delta k_j-\Delta k_l)\notag\\
	&+\pi+\theta(2k_j-2\eta)+O(1/N).
\end{align}
The formal expression $\rho(k,Q)=\frac{1}{L}\sum_{j=1}^{N}\delta(k-k_j)$ substituted into Eq.~(\ref{eq:rho}) gives
$1+\frac{1}{L}\sum_{j=1}^N \theta'(k_j-k)=2\pi \rho(k,Q)$. After introducing $g(k_j,Q)=L\rho(k_j,Q)\Delta k_j$, the latter equation enables us to express Eq.~(\ref{eq:Deltak}) as an integral equation
\begin{subequations}\label{eq:g}
\begin{gather}
	g(k,Q)-\frac{1}{2\pi}\int_{-Q}^{Q}dq \theta'(k-q)g(q,Q)=r(k,\eta),\\
	r(k,\eta)=\frac{1}{2}+ \frac{\theta(2k-2\eta)}{2\pi}.
\end{gather}
\end{subequations}
The energy of the system (\ref{eq:EP}) in the thermodynamic limit now becomes $E=N\epsilon_0+\mathcal{E}(Q,\eta)$, where $\epsilon_0$ is given by Eq.~(\ref{eq:E0}), while the energy of the polaron excitation corresponding to the momentum (\ref{eq:p}) is given by
\begin{align}\label{11}
\mathcal{E}(Q,\eta)=\frac{\hbar^2}{m}\int_{-Q}^{Q} dk\,\! k g(k,Q).
\end{align}
Here $g(k,Q)$ depends on $\eta$ and satisfies Eq.~(\ref{eq:g}).

In order to further transform Eq.~(\ref{11}), we introduce the Green's function for the linear operator in Eq.~(\ref{eq:g}) by \cite{reichert_exact_2019,takahashi}
\begin{align}\label{eq:green}
	G(k,k')-\frac{1}{2\pi}\int_{-Q}^{Q} dq\:\!\theta'(k-q)G(k',q)=\delta(k-k').
\end{align}
It is symmetric, $G(k,k')=G(k',k)$, as we can show by iterations. Multiplying Eq.~(\ref{eq:green}) by $r(k',\eta)$, after the integration over $k'$ we obtain Eq.~(\ref{eq:g}) provided $g(k,Q)=\int_{-Q}^{Q} dk' G(k,k') r(k',\eta)$. Equation~(\ref{11}) then becomes $\mathcal{E}(Q,\eta)=\int_{-Q}^{Q} dk\,\! \sigma(k,Q) r(k,\eta)$ where we have defined $\sigma(k,Q)=(\hbar^2/m)\int_{-Q}^{Q} dk' k' G(k,k')$. Multiplying Eq.~(\ref{eq:green}) by $k'$ and performing the integration over it, we obtain that $\sigma(k,Q)$ satisfies the integral equation
\begin{align}\label{eq:sigma}
	\sigma(k,Q)-\frac{1}{2\pi}\int_{-Q}^{Q} dq\:\!\theta'(k-q)\sigma(q,Q)=\frac{\hbar^2}{m}k.
\end{align}
Since $\sigma(k,Q)$ is an odd function of $k$ we eventually obtain
\begin{align}\label{eq:eps}
	\mathcal{E}(Q,\eta)=\frac{1}{2\pi}	\int_{-Q}^{Q} dk\:\! \sigma(k,Q)\theta(2k-2\eta).
\end{align}
Equation (\ref{eq:eps}) gives $\mathcal{E}=0$ at $\eta\to+\infty$, as expected.

Equations  (\ref{eq:p}) and (\ref{eq:eps}) are exact and determine the dispersion of the polaron excitation with the momentum $0\le p\le  2\pi\hbar n$ in the parametric form. It is difficult to find analytically the explicit form of the dispersion $\mathcal{E}(p)$  for arbitrary values of the interaction. Instead, here we study the dispersion at the smallest momenta. This is achieved if we expand $\theta(2k-2\eta)$ at $\eta/c\gg 1$ in Eqs.~(\ref{eq:p}) and (\ref{eq:eps}). Accounting for the leading-order term, we obtain the low-momentum spectrum  \cite{fuchs_spin_2005,zvonarev_edge_2009,Note2}
\footnotetext{Note that in the language of magnetic systems, the quadratic spectrum (\ref{eq:E(p)}) is expected for a long-wavelength spin-wave in ferromagnets.}
\begin{align}\label{eq:E(p)}
	\mathcal{E}(p)=\frac{p^2}{2m^*},
\end{align}
where the mass of the polaron excitation $m^*$ is given by
\begin{align}\label{eq:m*}
	\frac{1}{m^*}=\frac{S(Q)}{\pi \hbar^2 c\:\! n^2},\quad S(Q)=\int_{-Q}^{Q}dk\:\! k \sigma(k,Q).
\end{align}
Equation (\ref{eq:m*}) is the exact result.

The function $S(Q)$ [Eq.~(\ref{eq:m*})] and $\epsilon_0$ [Eq.~(\ref{eq:E0})] are not independent. We found that they satisfy a remarkable relation
\begin{align}\label{eq:s1}
S(Q)=2\pi n^2 \frac{\partial\:\! \epsilon_0(n)}{\partial n}.
\end{align}
Equation (\ref{eq:s1}) can be showed by differentiating Eq.~(\ref{eq:E0}) with respect to $n$ [which is a function of $Q$, see Eq.~(\ref{eq:n})], and then using the relation \cite{matveev_effective_2016}
\begin{align}\label{eq:ee}
\frac{\rho'_Q(k,Q)}{\rho(Q,Q)} +\frac{\sigma'_k(k,Q)}{\sigma(Q,Q)}= \frac{2\pi\hbar^2}{m\sigma(Q,Q)}\rho(k,Q)
\end{align}
between the functions $\rho(k,Q)$ and $\sigma(k,Q)$ satisfying Eqs.~(\ref{eq:rho}) and (\ref{eq:sigma}), respectively. Equation (\ref{eq:ee}) can be shown by differentiating Eq.~(\ref{eq:rho}) with respect to $Q$ and Eq.~(\ref{eq:sigma}) with respect to $k$, and then comparing the obtained expressions. From Eq.~(\ref{eq:s1}) we finally obtain 
\begin{align}\label{eq:1om*}
\frac{1}{m^*}=\frac{2}{\hbar^2 c} \frac{\partial\:\!\epsilon_0(n)}{\partial n},
\end{align}
which is our main result.

Equation (\ref{eq:1om*}) is exact and thus valid at arbitrary interaction $c$. It shows that the polaron mass in the Yang-Gaudin model depends only on the derivative with respect to the density of the ground-state energy per particle of the single-component Lieb-Liniger  Bose gas (\ref{eq:E0}). The latter can be expressed in terms of the dimensionless function $e(\gamma)$ defined by \cite{lieb_exact_1963}
\begin{align}\label{eq:E0LL}
	\epsilon_0=\frac{\hbar^2 n^2}{2m}  e(\gamma).
\end{align}
Here $\gamma=c/n$ is the dimensionless parameter that accounts for the interaction strength. The particular form (\ref{eq:E0LL}) enables us to express the polaron mass as
\begin{align}\label{eq:m*LL}
	\frac{m}{m^*}=-\gamma^2 \frac{\partial}{\partial\gamma}\left(\frac{e(\gamma)}{\gamma^2}\right),
\end{align}
which is an alternative form of Eq.~(\ref{eq:1om*}). Equations (\ref{eq:1om*}) and (\ref{eq:m*LL}) apply in the thermodynamic limit.

\begin{figure}
	\includegraphics[width=\columnwidth]{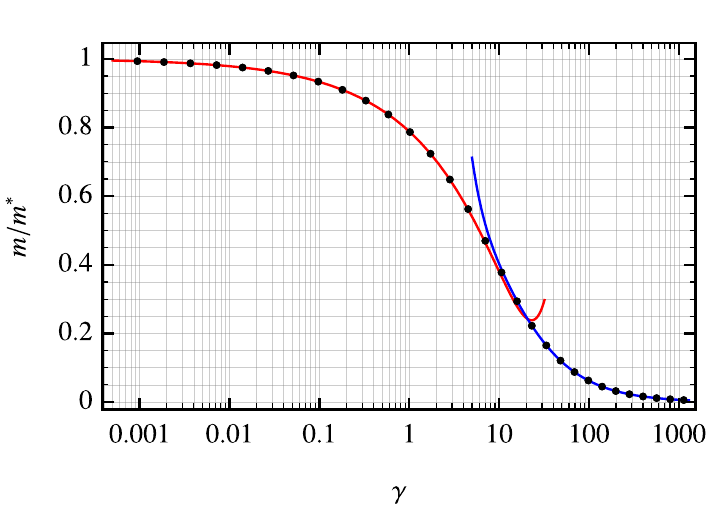}
	\caption{Plot of the inverse of the polaron mass as a function of the interaction strength $\gamma$. The dots represent the exact result obtained directly from Eq.~(\ref{eq:m*}), while the two curves are the analytical results at small and large $\gamma$ given by Eqs.~(\ref{eq:m*gamma<<1}) and (\ref{eq:m*gamma>>1}), respectively.} \label{fig1}
\end{figure}

The function $e(\gamma)$ defined by Eq.~(\ref{eq:E0LL}) can be routinely evaluated numerically \cite{lieb_exact_1963,ristivojevic_conjectures_2019}. On the other hand, to date there is no simple closed expression for $e(\gamma)$. It was calculated perturbatively at $\gamma\ll 1$ and $\gamma\gg 1$ at low orders \cite{lieb_exact_1963} and recently to very high orders \cite{ristivojevic_excitation_2014,ristivojevic_conjectures_2019,marino_exact_2019}. From $e(\gamma)$ at weak interaction calculated in Refs.~\cite{ristivojevic_conjectures_2019,marino_exact_2019} we obtain the power series
\begin{align}\label{eq:m*gamma<<1}
\frac{m}{m^*}={}&
1-\frac{2}{3\pi}\sqrt\gamma+\frac{4-3\zeta(3)}{16\pi^3}\gamma^{3/2}+				\frac{4-3\zeta(3)}{24\pi^4}\gamma^{2}\notag\\
&+\frac{3[32-60\zeta(3)+45\zeta(5)]}{2048\pi^5}\gamma^{5/2} +O(\gamma^3)
\end{align}
in the regime $\gamma\ll 1$. Equation (\ref{eq:m*gamma<<1}) shows that decreasing the interaction strength, the polaron mass approaches the bare impurity mass. This is expected, since the impurity in this limit becomes decoupled from the system. We notice the absence of the term proportional to $\gamma$ in Eq.~(\ref{eq:m*gamma<<1}), which is obvious from the form of the derivative in Eq.~(\ref{eq:m*LL}). The first two terms of the expression for $m/m^*$ were calculated in Ref.~\cite{batchelor_collective_2006} using the discrete Bethe ansatz and in Ref.~\cite{fuchs_spin_2005} using the hydrodynamic description \cite{Note3}.\footnotetext{Note that the first two term in Eq.~(\ref{eq:m*gamma<<1}) follow from Eq.~(D15) of Ref.~\cite{reichert_quasiparticle_2017} taken at $M=m$, $G=g$, and $k_F=0$.}\nocite{reichert_quasiparticle_2017} Our expression (\ref{eq:m*gamma<<1}) that relies on the Bethe ansatz is in agreement with Ref.~\cite{fuchs_spin_2005}. However, we calculated several further terms of the series, which would be very hard to obtain using alternative methods. 

The function $e(\gamma)$  at strong interaction was calculated in Ref.~\cite{ristivojevic_excitation_2014}. It enables us to find
\begin{align}\label{eq:m*gamma>>1}
\frac{m}{m^*}=\frac{2\pi^2}{3\gamma}-\frac{4\pi^2}{\gamma^2}+\frac{16\pi^2}{\gamma^3}- \frac{32\pi^2(15-\pi^2)}{9\gamma^4}+O(\gamma^{-5})	
\end{align}
in the regime $\gamma\gg 1$. At strong interaction, the impurity motion implies the motion of many surrounding bosons. The resulting polaron quasiparticle is therefore heavy due to the surrounding cloud; its effective mass is determined by Eq.~(\ref{eq:m*gamma>>1}). The leading-order term in Eq.~(\ref{eq:m*gamma>>1}) was known before \cite{fuchs_spin_2005,batchelor_collective_2006}. We note that the first subleading term was calculated in Ref.~\cite{batchelor_collective_2006}, but it disagrees within a numerical coefficient from our result (\ref{eq:m*gamma>>1}).  Plotted in Fig.~\ref{fig1} is  the exact result for the polaron mass obtained by numerically solving the integral equations (\ref{eq:rho}) and (\ref{eq:sigma}) and its comparison with the two analytical expressions (\ref{eq:m*gamma<<1}) and (\ref{eq:m*gamma>>1}).

In conclusion, analyzing the Bethe ansatz equations of the Yang-Gaudin model for a Bose gas, we have found the exact result for the polaron mass $m^*$. It is given by Eq.~(\ref{eq:1om*}) or its equivalent form (\ref{eq:m*LL}). The latter expressions show that $m^*$ is fully determined by the dependence of the ground-state energy per particle of the Lieb-Liniger Bose gas on density or equivalently on $\gamma$. The latter function, and thus $m^*$, can be calculated analytically at very high orders at weak or strong interaction \cite{ristivojevic_excitation_2014,marino_exact_2019}. We note that the specific form of the scattering phase shift is not used in Eqs.~(\ref{eq:rho})--(\ref{eq:eps}) and~(\ref{eq:ee}). Therefore, the described procedure might be applicable in studies of other two-component models \cite{sutherland} that can also describe polarons.

Useful discussions with B. Reichert are gratefully acknowledged. 

%\bibliography{bibliography.bib}


\end{document}